# Large-Scale Paralleled Sparse Principal Component Analysis

W. Liu[1], H. Zhang[1], D. Tao[1*], Y. Wang[1], K. Lu[1]

[1] China University of Petroleum (dtao.scut@gmail.com)

**Abstract**

Principal component analysis (PCA) is a statistical technique commonly used in multivariate data analysis. However, PCA can be difficult to interpret and explain since the principal components (PCs) are linear combinations of the original variables. Sparse PCA (SPCA) aims to balance statistical fidelity and interpretability by approximating sparse PCs whose projections capture the maximal variance of original data. In this paper we present an efficient and paralleled method of SPCA using graphics processing units (GPUs), which can process large blocks of data in parallel. Specifically, we construct parallel implementations of the four optimization formulations of the generalized power method of SPCA (GP-SPCA), one of the most efficient and effective SPCA approaches, on a GPU. The parallel GPU implementation of GP-SPCA (using CUBLAS) is up to eleven times faster than the corresponding CPU implementation (using CBLAS), and up to 107 times faster than a MatLab implementation. Extensive comparative experiments in several real-world datasets confirm that SPCA offers a practical advantage.

## 1. Introduction

Principal component analysis (PCA) [1] is a well-established tool used for data analysis and dimensionality reduction. The goal of PCA is to find a sequence of orthogonal factors that represent the directions of largest variance. PCA is used in many applications, including machine learning, image processing, neurocomputing, engineering, and computer networks, especially for large datasets. However, despite its power and popularity, a major limitation of PCA is that the derived principal components (PCs) are difficult to interpret and explain because they tend to be linear combinations of all the original variables.

Over the past ten years, sparse principal component analysis (SPCA) has been used to improve the interpretability of PCs. SPCA aims to find a reasonable balance between statistical fidelity and interpretability by approximating sparse PCs. Briefly, SPCA methods can be divided into two groups: (1) ad hoc methods [2] [3] and (2) sparsity penalization methods [4] [5] [6] [7] [8] [9] . Ad hoc methods post-process the components obtained from classical PCA; for example, Jolliffe [2] uses rotation techniques in the standard PCA subspace to find sparse loading vectors, while Cadima and Jolliffe [3] simply set the PCA loadings with small absolute values to zero. Sparsity penalization methods usually formulate the SPCA problem as an optimization program by adding a sparsity-penalized term into the PCA framework. For example, Jolliffe et al.[4] maximize the Rayleigh quotient of the data covariance matrix under the L1-norm penalty in the SCoTLASS algorithm. Zou et al.[5]

formulate sparse PCA as a regression-type optimization problem by imposing the LASSO penalty on the regression coefficients. In the DSPCA algorithm, d'Aspremont et al. [6] solve a convex relaxation of the sparse PCA, while Moghaddam et al.[8] and d'Aspremont et al.[7] go on to use greedy methods in order to solve the combinatorial problems encountered in sparse PCA. Finally, Journée et al.[9] propose the generalized power method for sparse PCA (GP-SPCA), in which sparse PCA is formulated as two single-unit and two block optimization problems. GP-SPCA has optimal convergence properties when either the objective function, or the feasible set, are strongly convex [9] .

There is ever growing collection, sharing, combination, and use of massive amounts of data. The analysis of such "big data" has become essential in many commercial and scientific applications, from image analysis to genome sequencing. Parallel computing algorithms are essential for large-scale, high-dimensional data. Fortunately, modern graphics processing units (GPUs) have a highly parallel structure that makes them ideally suited to processing big data algorithms as well as graphics [10] .

In this study we consider how to build compact, unsupervised representations of large-scale, high-dimensional data using sparse PCA schemes, with an emphasis on executing the algorithm in the GPU environment. The work can be regarded as a set of parallel optimization procedures for SPCA; specifically, we construct parallel implementations of the four optimization formulations used in GP-SPCA. To the best of our knowledge, GP-SPCA has not previously been implemented using GPUs. We compare the GPU implementation (on an NVIDIA Tesla C2050) with the corresponding CPU implementation (on a six-core 3.33 GHz high-performance cluster) and show that the parallel GPU implementation of GP-SPCA is up to 11 times faster than the corresponding CPU implementation, and up to 107 times faster than the corresponding MatLab implementation. We also conduct extensive comparative experiments of SPCA and PCA on several benchmark datasets, which provide further evidence that SPCA outperforms PCA in the majority of cases.

The remainder of this paper is organized as follows. GP-SPCA is briefly introduced in Section 2. The implementation of GP-SPCA on GPUs using CUBLAS is described in Section 3, and the experiments are presented in Section 4. We conclude in Section 5.

## 2. Generalized power method of SPCA

Let $A \in R^{p \times n}$ be a matrix encoding $p$ samples of $n$ variables. SPCA aims to find principal components that are both sparse and explain as much of the variance in the data as possible, and in doing so finds a reasonable trade-off between statistical fidelity and interpretability. GP-SPCA considers two single-unit and two block formulations of SPCA, in order to extract m sparse principal components, with $m = 1$ for two single-unit formulations of SPCA and $p \geq m \geq 1$ for the two block formulations of SPCA. GP-SPCA maximizes a convex function on the unit Euclidean sphere in $R^p$ (for $m = 1$) or on the Stiefel manifold in $R^{p \times m}$ (for $m > 1$).

Depending on the type of penalty (either $l_1$ or $l_0$) used to enforce sparsity, there are four formulations of SPCA, namely single-unit SPCA via the $l_1$-penalty (GP-SPCA-SL1), single-unit SPCA via the $l_0$-penalty (GP-SPCA-SL0), block SPCA via the $l_1$-penalty (GP-SPCA-BL1), and block SPCA via the $l_0$-penalty (GP-SPCA-BL0).

Denote the unit Euclidean ball (resp. sphere) in $R^k$ by $B^k = \{y \in R^k | y^T y \leq 1\}$ (resp. $S^k = \{y \in R^k | y^T y = 1\}$). Denote the space of $n \times m$ matrices with unit-norm columns by $[S^n]^m = \{Y \in R^{n \times m} | Diag(Y^T Y) = I_m\}$, where $Diag(\cdot)$ is the diagonal matrix, by extracting the diagonal of the argument. Denote the Stiefel manifold by $S_m^p = \{Y \in R^{n \times m} | Y^T Y = I_m\}$, and write $sign(t)$ for the sign of the argument $t \in R$ and $t_+ = max\{0, t\}$. The characteristics of the four variants are summarized in Table 1.

Table 1. The four variant formulations of GP-SPCA.

| | **Original form of SPCA** | **Reformulation** |
|---|---|---|
| **GP-SPCA-SL1** | $\phi_{l_1}(\gamma) \equiv \max\limits_{z \in B^n} \sqrt{z^T \Sigma z} - \gamma \|z\|_1$ | $\phi_{l_1}^2(\gamma) \equiv \max\limits_{x \in S^p} \sum\limits_{i=1}^{n} [|a_i^T x| - \gamma]_+^2$ |
| **GP-SPCA-SL0** | $\phi_{l_0}(\gamma) \equiv \max\limits_{z \in B^n} z^T \Sigma z - \gamma \|z\|_0$ | $\phi_{l_0}(\gamma) \equiv \max\limits_{x \in S^p} \sum\limits_{i=1}^{n} [(a_i^T x)^2 - \gamma]_+$ |
| **GP-SPCA-BL1** | $\phi_{l_{1,m}}(\gamma) \equiv \max\limits_{\substack{X \in S_m^p \\ Z \in [S^n]^m}} Tr(X^T AZN) - \sum\limits_{j=1}^{m} \gamma_j \sum\limits_{i=1}^{n} |z_{ij}|$ | $\phi_{l_{1,m}}^2(\gamma) \equiv \max\limits_{X \in S_m^p} \sum\limits_{j=1}^{m} \sum\limits_{i=1}^{n} [\mu_j |a_i^T x_j| - \gamma_j]_+$ |
| **GP-SPCA-BL0** | $\phi_{l_{0,m}}(\gamma) \equiv \max\limits_{\substack{X \in S_m^p \\ Z \in [S^n]^m}} Tr(Diag(X^T AZN)^2) - \sum\limits_{j=1}^{m} \gamma_j \|z_j\|_0$ | $\phi_{l_{0,m}}(\gamma) \equiv \max\limits_{X \in S_m^p} \sum\limits_{j=1}^{m} \sum\limits_{i=1}^{n} [(\mu_j a_i^T x)^2 - \gamma_j]_+$ |

GP-SPCA has optimal convergence properties when either the objective functions, or the feasible set, are strongly convex, which is the case with the single-unit formulations and can be enforced in the block cases [9].

## 3. GPU implementation of GP-SPCA

GPUs are typically used for computer graphics processing in general-purpose computing. There is a discrepancy between the floating-point capability of the CPU and GPU because the GPU is specialized for intensive, highly-parallel computation, and is therefore specifically designed to devote more transistors to data processing rather than data caching and flow control, as shown in Figure 1[10].

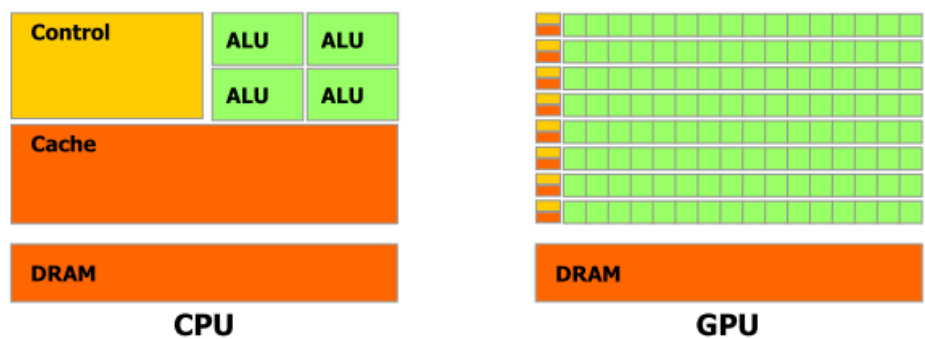

Figure 1. The difference between GPU and CPU [10]

$CUDA^{TM}$ is a general-purpose parallel computing architecture designed by NVIDIA, which has a parallel programming model and instruction set architecture. CUDA guides the programmer to partition a problem into a sub-problem that can be solved as independent parallel blocks of threads in a thread hierarchy; Figure 2 illustrates the hierarchy of threads, blocks, and grids used in CUDA. As well as the CUDA programming environment, NVIDIA also supplies toolkits for the programmer: CUBLAS [11] is one such library that implements Basic Linear Algebra Subprograms (BLAS).

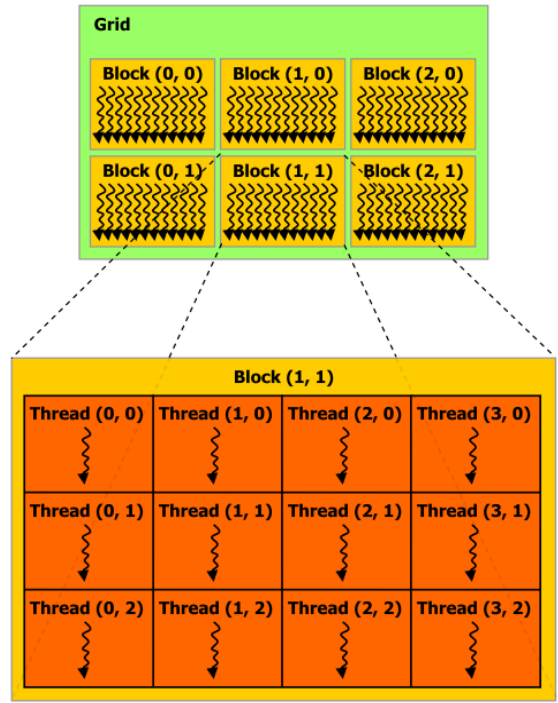

Figure 2. Grids of thread blocks[10]

Here we implement all formulations of GP-SPCA on the GPU using CUBLAS. The data space is allocated both on the host memory (CPU) and on the device memory (GPU). Data are initialized on the host memory before being transferred to the device memory, after which parallel computation is performed on the device memory. The results are then transferred back to the host memory when computation is complete.

## 4. Experiments

In this section, we conduct comparative experiments to evaluate the efficiency of GPU computing and the effectiveness of GP-SPCA.

### 4.1 Efficiency of GPU computing

In order to compare the efficiency of GPU and CPU computing, we first conduct the

CPU implementation of GP-SPCA using GSL CBLAS [12], which is a highly efficient implementation of BLAS. We also compare the implementation with the MatLab application presented in [9].

A six-core 3.33 GHz high performance cluster was used for the CPU implementation, and an NVIDIA Tesla C2050 for the GPU implementation. Twenty test instances were generated for each input matrix $A_{P\times N}$ ($N \in [5.0 \times 10^2, 3.2 \times 10^4]$, $P = N/10$). Here, $m = 5$ is the number of sparse PCs, and $\gamma \in \{0.01, 0.05\}$ is the aforementioned parameter that balances the sparsity and variance of the PCs.

Figure 3 shows the average running time of different input matrices using different parameters. The x-axis indicates the size of the input matrix and the y-axis denotes computation time. The difference in processing time (between CPU and GPU) increases with increasing size of the input matrix, with up to eleven times improvement in speed over the corresponding CBLAS implementation, and up to 107-times over the MatLab implementation.

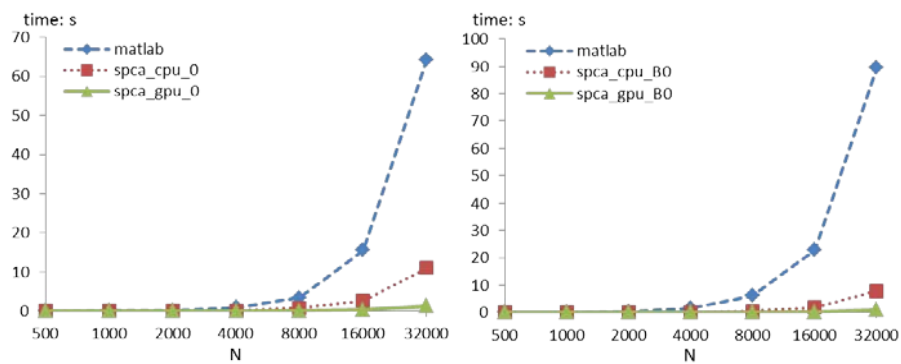

a. GP-SPCA-SL0, $m = 5$, $\gamma = 0.01$    b. GP-SPCA-BL0, $m = 5$, $\gamma = 0.01$

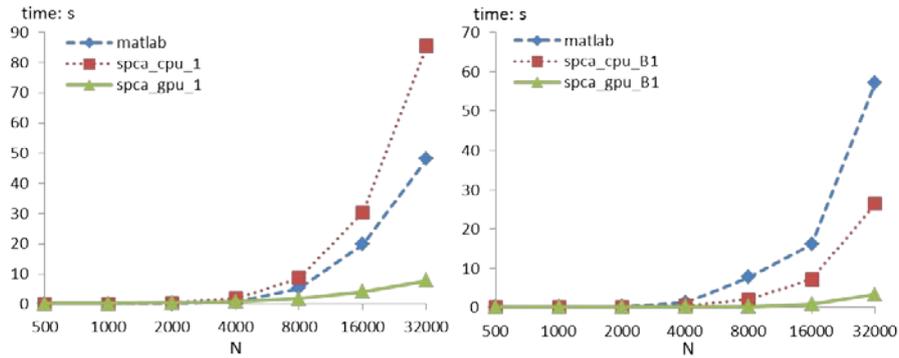

c. GP-SPCA-SL1, $m = 5, \gamma = 0.05$     d. GP-SPCA-BL1, $m = 5, \gamma = 0.05$

Figure 3. A comparison of GP-SPCA performed on a GPU (Tesla C2050) and a CPU

**4.2 Effectiveness of GP-SPCA**

To evaluate the effectiveness of GP-SPCA in practice, we next conducted GP-SPCA and PCA experiments on several benchmark datasets, including the USPS database [13] , the COIL20 database [14] , and the Isolet spoken letter recognition database [15] . For each experiment, we used GP-SPCA and PCA to learn the project functions using training samples, before mapping all the samples (both training and test samples) into the lower dimensional subspace where recognition is performed using a nearest neighbor classifier.

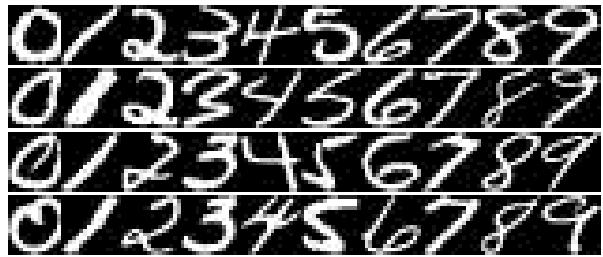

Figure 4. Examples of handwriting in the USPS database

**USPS database:**

The USPS database [13] is a handwritten digit database containing 9298 16×16 pixel handwritten digit images in total (Figure 4). The database was split into 7291 training images and 2007 test images as in [16] [17] , with the parameter $\gamma$ set to 0.1.

The results of SPCA and PCA in recognizing the ten handwritten digits are shown in Figure 5, from which we can see that SPCA outperforms PCA in most cases.

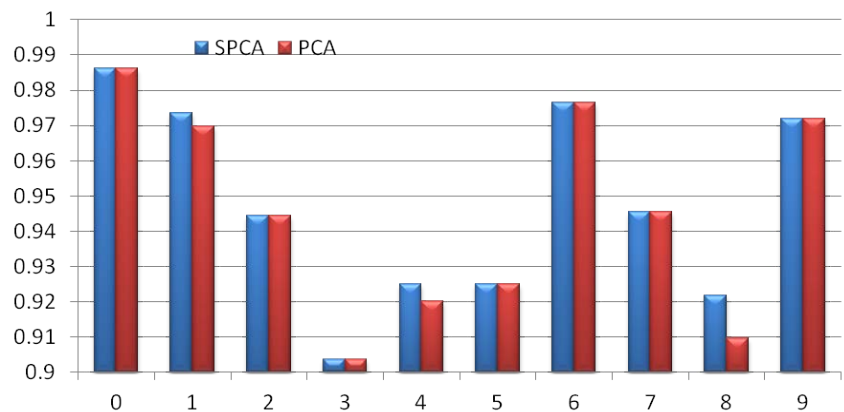

Figure 5. Recognition of SPCA and PCA on USPS

**COIL20 database:**

The COIL20 database [14] contains 1440 images of 20 objects (for examples, see Figure 6). The images of each object are taken five degrees apart as the object is rotated on a turntable, and as a result each object is represented by 72 $32 \times 32$ pixel images. We randomly selected two groups of 24 and 36 examples of each object as training sets, and used the remaining images for the test sets. The parameter $\gamma$ was set to 0.3 for 24-example group, and 0.1 for the 36-example group. All the experiments were repeated five times.

Figure 7 shows that SPCA outperforms PCA in both cases. Figure 8, which shows the recognition rate of selected objects, demonstrates that SPCA outperforms PCA in most cases.

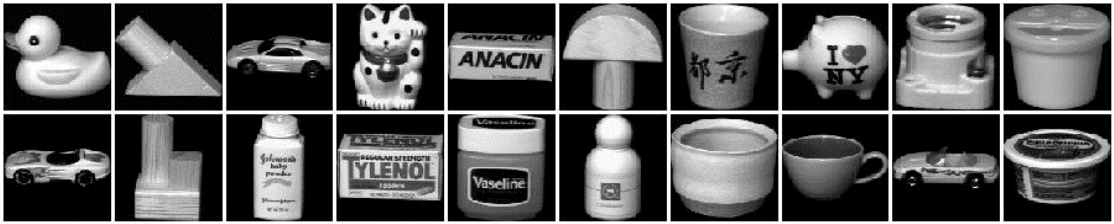

Figure 6. COIL20 examples

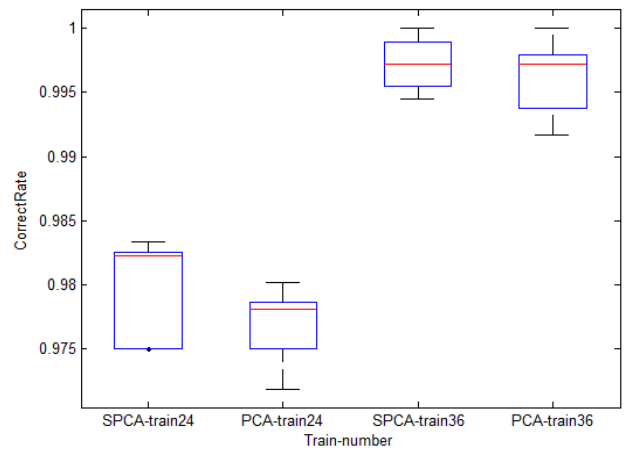

Figure 7. The average recognition rates of SPCA and PCA on COIL20 data

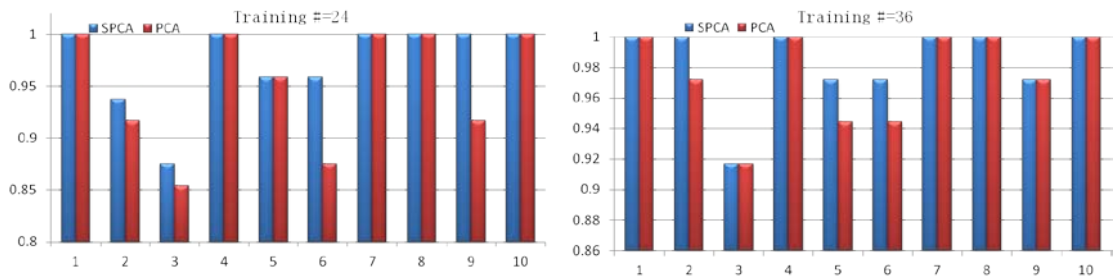

Figure 8. The recognition results of selected objects

**Isolet spoken letter recognition database:**

The Isolet spoken letter recognition database [15] contains 150 subjects, each of whom speaks each letter of the alphabet twice. The speakers were grouped into five sets of 30 speakers; three were used for training and two for testing in the first experiment and four groups for training the other for testing in the second experiment

(to evaluate robustness). The parameter $\gamma$ was set to $10^{-6}$ for the first experiment and 0.02 for the second, and each experiment was repeated five times.

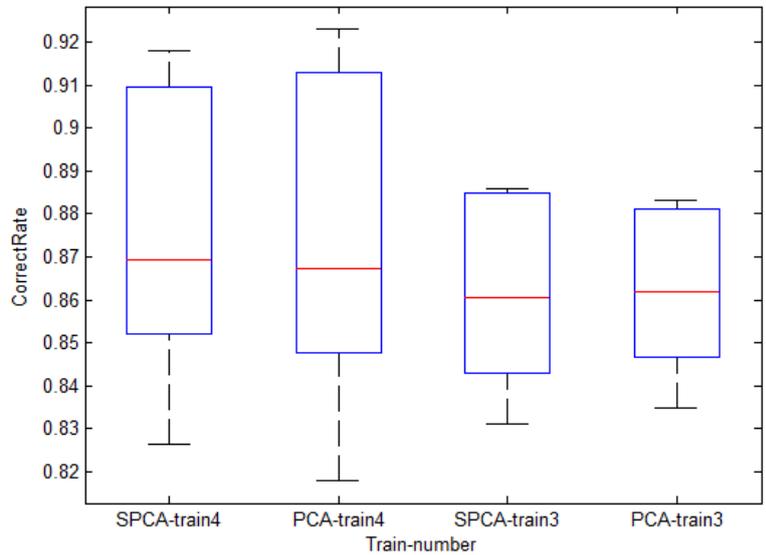

Figure 9. The average recognition rates of SPCA and PCA on Isolet data

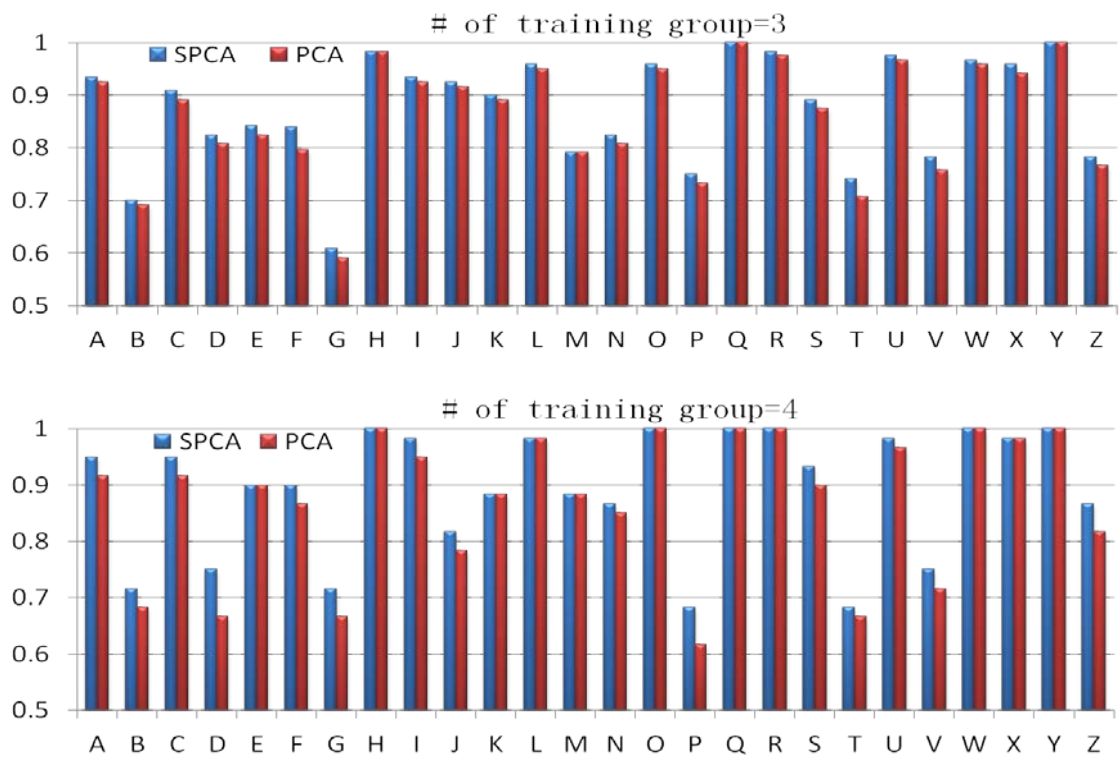

Figure 10. Recognition rates for each character

Figures 9 and 10 show the average recognition rates and recognition of each character,

respectively. SPCA is superior to PCA in the majority of cases.

5. Conclusion

Sparse PCA is a reasonable method for balancing statistical fidelity and interpretability. In this paper, we present a paralleled method of GP-SPCA, one of the most efficient SPCA approaches, using a GPU. Specifically, we construct parallel implementations of the four optimization formulations for the GPU, and compare this with a CPU implementation using CBLAS. Using real-world data, we experimentally validate the effectiveness of GP-SPCA and demonstrate that the parallel GPU implementation of GP-SPCA can significantly improve performance. This work has several potential applications in large-scale, high-dimension reduction problems such as video indexing[18] [21] and web image annotation[19] [20] , which will be the subject of future study.